\documentstyle[11pt,twoside,jltp,epsfig]{article}

\title{Magnetic Interaction in Insulating Cuprates}

\author{Yoshiaki Mizuno, Takami Tohyama and Sadamichi Maekawa\address{Institute for Materials Research, Tohoku University, Sendai 980-8577, Japan}}
\runninghead{Y. Mizuno, T. Tohyama and S. Maekawa}{Magnetic Interaction in Insulating Cuprates}
\begin{document}
\begin{abstract}
We examine two-spin (2S) and cyclic four-spin (4S) magnetic interactions in insulating ladder and two-dimensional (2D) cuprates. By a comparison of eigenstates between $d$-$p$ and Heisenberg models, we evaluate magnitudes of these interactions. We find that the 4S interaction is $\sim$10~$\%$ of nearest neighboring 2S interaction, and a diagonal 2S interaction is considerably small. The 4S interaction for a ladder cuprate is larger than that for 2D one, and plays an important role in the low-energy excitation. The Heisenberg ladder with the obtained 2S and 4S interactions reproduces very well the experimental result of the temperature dependence of the magnetic susceptibility.

\end{abstract}

\maketitle

%Include this space if you do not use sections in your document.
\vspace{0.3in}

A variety of insulating cuprates, including the parent compounds of high-${\it T}$$_c$ superconductors, afford us an opportunity to study magnetic properties of low-dimensional systems. They have unique Cu-O structures formed by CuO$_4$ tetragons, that is, two dimensional (2D) CuO$_2$ planes, two-leg ladders and one dimensional (1D) chains. In insulating cuprates, magnetic interactions $J$ between Cu S=1/2 spins become important key parameters for understanding of the electronic states. In 180$^{\circ}$ Cu-O-Cu coupling, which is realized when CuO$_4$ tetragons share their corners, the superexchange process via O ions gives a dominant contribution to $J$.

Recent magnetic measurements demonstrate remarkable dependence of $J$ between nearest neighboring (NN) Cu spins on the dimensionality of Cu-O structure:\cite{Mizuno} (i) $J$ in 1D cuprates is larger than that in 2D ones, and (ii) for ladder cuprates, $J$ along leg direction is larger than that along rung one. These can not be understood only by considering the bond length dependence of $J$. In our previous study,\cite{Mizuno} we have shown that the hopping matrix elements between Cu3$d$ and O2$p$ orbitals ($T_{pd}$) and between O2$p$ ones ($T_{pp}$) are considerably influenced by Madelung potentials around Cu-O and O-O bonds. As a result, $J$ depends on the crystal structure through the modification of $T_{pd}$ and $T_{pp}$ due to the Madelung potential. By taking into account these effects, we have clarified the microscopic origin of the characteristics of (i) and (ii). 

Very recently, an inelastic neutron experiment for the two-leg ladder cuprate La$_6$Ca$_8$Cu$_{24}$O$_{41}$ has been performed in the wide range of the excitation energy.\cite{Matsuda} By comparing the observed dispersion of spin excitation with the theoretical one, it has been shown that $J$ along leg direction is larger that along rung one, and in addition a four-spin (4S) interaction ($\sim$0.005~eV) is necessary to explain the experiment. Here, the 4S interaction is given by the cyclic permutation on four Cu spins forming a plaquette (see Fig.~1(a-2)). This result indicates importance of not only usual two-spin (2S) interactions but also 4S interaction among Cu spins in a plaquette. Therefore, it is necessary to establish a proper magnetic description for undoped cuprates. This will also yield an appropriate understanding of magnetic properties in high-$T_{\it c}$ cuprates.

So far, some theoretical studies on the 4S interaction have been done, but they are only for 2D cuprates\cite{Roger,Schmidt}. In this paper, focusing on insulating ladder and 2D cuprates, we investigate possible magnetic interactions including 4S one in the systems. By associating a $d$-$p$ model with an effective Heisenberg model, we evaluate magnitudes of these magnetic interactions. The effect of 4S interaction on the magnetic properties is examined through the magnetic susceptibility, and is shown to lead to a dramatic change in the low-energy excitation for ladder cuprates.

As a starting model for a ladder cuprate, we consider a $d$-$p$ model as shown in Fig.~1(a-1), where three and two Cu3$d_{x^2-y^2}$ orbitals are arranged along leg ($x$) and rung ($y$) directions, respectively, and O2$p_{\sigma}$ orbitals ($\sigma=x,y$) form the CuO$_4$ tetragons (Cu$_6$O$_{17}$ cluster). In the model, open boundary condition is imposed. The Hamiltonian to describe the model is expressed by $T_{pd}$ and $T_{pp}$, the energy-level separation between Cu3$d$ and O2$p$ orbitals ($\Delta$) and Coulomb interactions such as on-site Coulomb interactions at Cu site ($U_d$) and O one ($U_p$).

An effective model to describe the low-energy excitation of the $d$-$p$ model can be given as a S=1/2 Heisenberg model with 2S interactions along leg ($J_{\rm leg}$), rung ($J_{\rm rung}$) and diagonal ($J_{\rm diag}$) and cyclic 4S interaction ($J_{\rm cyc}$) as shown in Fig.~1(a-2). The spin Hamiltonian is written by 
\begin{eqnarray}
H&=&\sum_{\bf i} (J_{\rm leg} {\bf S}_{\bf i} \cdot {\bf S}_{{\bf i}+\hat{\bf x}} + J_{\rm rung} {\bf S}_{\bf i} \cdot {\bf S}_{{\bf i}+\hat{\bf y}} + J_{\rm diag} {\bf S}_{\bf i} \cdot {\bf S}_{{\bf i}+\hat{\bf x}+\hat{\bf y}}) \nonumber \\
&+& J_{\rm cyc} \sum_{\rm plaquette}(P_{\bf ijkl}+P_{\bf ijkl}^{-1}),
\end{eqnarray}
where ${\bf S}_{\bf i}$ is a spin operator at {\bf i} site, and $\hat{\bf x}$ ($\hat{\bf y}$) is a unit vector along $x$ ($y$)  directions. Here, $J_{\rm cyc}$ is defined as a coefficient of 4S cyclic permutation operators $P_{\bf ijkl}$ and $P_{\bf ijkl}^{-1}$, which can be rewritten by using the 2S and 4S interactions as
\begin{eqnarray}
P_{\bf ijkl}+P_{\bf ijkl}^{-1}=4[({\bf S}_{\bf i} \cdot {\bf S}_{\bf j})({\bf S}_{\bf k} \cdot {\bf S}_{\bf l})+({\bf S}_{\bf i} \cdot {\bf S}_{\bf l})({\bf S}_{\bf j} \cdot {\bf S}_{\bf k})-({\bf S}_{\bf i} \cdot {\bf S}_{\bf k})({\bf S}_{\bf j} \cdot {\bf S}_{\bf l})] \nonumber \\
+[({\bf S}_{\bf i} \cdot {\bf S}_{\bf j})+({\bf S}_{\bf j} \cdot {\bf S}_{\bf k})+({\bf S}_{\bf k} \cdot {\bf S}_{\bf l})+({\bf S}_{\bf l} \cdot {\bf S}_{\bf i})]+[({\bf S}_{\bf i} \cdot {\bf S}_{\bf k})+({\bf S}_{\bf j} \cdot {\bf S}_{\bf l})]+\frac{1}{4}.
\end{eqnarray}

We calculate the eigenstates and eigenvalues for the $d$-$p$ model by numerically diagonalization method. For the ladder cuprate, SrCu$_2$O$_3$, $T_{pd}$'s are 1.08, 1.22 and 1.25~eV for leg, rung and interladder directions, respectively. $\Delta$=2.6~eV, $U_d$=8.5~eV, $U_p$=4.1~eV and the direct exchange interaction between Cu3$d$ and O2$p$ orbitals ($K_{pd}$) is 0.05~eV. $T_{pp}$'s are 0.52~eV and 0.42~eV for solid and dotted  bonds in Fig.~1(a), respectively.\cite{Mizuno} Note that the magnitudes of two kind of $T_{pp}$'s are different because of a difference of local environment,\cite{Mizuno} $i.e.$ the former has only one neighbor Cu sites,  while the latter does two neighbor Cu sites, where one of them belongs to the neighboring ladder. 

By associating the obtained eigenstates and eigenvalues of the $d$-$p$ model with those of the Heisenberg model of Eq.~(1), we determine $J_{\rm leg}$, $J_{\rm rung}$, $J_{\rm diag}$ and $J_{\rm cyc}$. The five energy differences obtained by six lowest eigenvalues of the $d$-$p$ model are used to fix these interactions.  In addition to Cu$_6$O$_{17}$ cluster, we also used a quadratic Cu$_4$O$_{12}$ cluster to determine the magnetic interactions for the 2D cuprate, La$_2$CuO$_4$ (see Ref.~1 for the parameters used in the calculation), as well as ladder cuprates.

\begin{table}[h]
\caption{The magnetic interactions (energy unit in eV) for Cu$_4$O$_{12}$ cluster (A) and Cu$_6$O$_{17}$ cluster (B)  obtained by the fit to Heisenberg models. The numbers in parentheses represent the deviation in the last significant digit.}
\begin{tabular}{c|ccc|cccc}
\hline
\hline
&\multicolumn{3}{c|}{\rm 2D cuprate (La$_2$CuO$_4$)}&\multicolumn{4}{c}{\rm Ladder cuprate (SrCu$_2$O$_3$)}\\
%\cline{2-8}
&$J_{\rm NN}$&$J_{\rm diag}$&$J_{\rm cyc}$&$J_{\rm leg}$&$J_{\rm rung}$&$J_{\rm diag}$&$J_{\rm cyc}$\\
\hline
(A) & 0.146(1)& 0.000(0) & 0.0108(8) & 0.165(5) & 0.15(0) & 0.001(1) & 0.015(0) \\
(B) & & & & 0.195(5) & 0.15(2) & 0.003(2) & 0.018(2)\\
\hline
\hline
\end{tabular}
\end{table}

The results are summarized in Table~1. The results for a 2D cuprate are comparable with those by Schmidt {\it et al.}.\cite{Schmidt}   $J_{\rm cyc}$ is $\sim$7 \% of the 2S NN interaction $J_{\rm NN}$, and $J_{\rm diag}$ is zero. $J_{\rm NN}$ is $\sim$ 0.15~eV, similar to our previous study.\cite{Mizuno} As for a ladder cuprate, we find that $J_{\rm leg}>J_{\rm rung}$ in both Cu$_4$O$_{12}$ and Cu$_6$O$_{17}$ clusters, which is again  consistent with our previous study.\cite{Mizuno} The anisotropy is caused by the enhancement of interladder $T_{pp}$ denoted by solid line in Fig.~1(a-1). In addition, the enhancement enlarges the value of $J_{\rm cyc}$ for a ladder cuprate as compared with that for a 2D cuprate, and its magnitude is $\sim$10 \% of $J_{\rm leg}$. $J_{\rm diag}$ is almost zero. The energy levels of the Heisenberg model are shown in Fig.~1 (b) together with those for the $d$-$p$ model, where all the eigenvalues $E_i$ ($i$=0$\sim$6) of the Heisenberg model is shifted so that the energy of ground state $E_0$ is equal to that of the $d$-$p$ model. We can see a good agreement of the energy levels and eigenstates between both models. In order to check validity of the obtained interactions, we have also performed calculations for a $d$-$p$ model with periodic boundary condition (Cu$_6$O$_{15}$ cluster) and the corresponding periodic Heisenberg model. We confirmed that the Heisenberg model with the interactions in Table~1 reproduces very well the distribution of the eigenstates for the periodic $d$-$p$ model.

By using the magnetic interactions obtained, we examine the temperature dependence of the  magnetic susceptibility $\chi(T)$. The exact diagonalization method is used to get the $\chi(T)$'s for Heisenberg ladders with 2$\times$4, 2$\times$6 and 2$\times$8 sites. The results are shown in Fig.~3 for the parameters of $J_{\rm leg}$=0.19~eV, $J_{\rm rung}$=0.15~eV, $J_{\rm diag}$=0.003~eV and $J_{\rm cyc}$=0.0175~eV and of the same values but $J_{\rm cyc}$=0~eV. In the inset, $\chi(T)$'s at wider range of temperature are shown.

The effects of $J_{\rm cyc}$ on $\chi(T)$ are (i) to enhance the magnitude of $\chi(T)$ at low temperatures and (ii) to shift a hump structure ($\sim$2000~K) in the lower temperature region. This suggests the decrease of an energy gap between the singlet and triplet states (spin gap). The decrease of the spin gap due to $J_{\rm cyc}$ is also observed in the dynamical spin correlation function.\cite{Brehmer} The reason for the decrease is that $J_{\rm cyc}$ disturbs the singlet states on rungs as magnetic frustration, consequently making $J_{\rm rung}$ effectively decreased. We find that by considering $J_{\rm cyc}$, $\chi(T)$ is in agreement with the experimental $\chi(T)$ for SrCu$_2$O$_3$\cite{Azuma}. Thus, it is interesting that $J_{\rm cyc}$ strongly influences the low-energy excitation above the spin gap. This is in contrast to 2D cuprates, where $J_{\rm cyc}$ only modifies the values of 2S interactions. Further study on the effect of $J_{\rm cyc}$ on other physical quantities is necessary for a quantitative investigation between theory and experiment.

This work was supported by CREST and NEDO. The parts of the numerical calculation were performed in the Supercomputer Center in ISSP, Univ. of Tokyo, and the supercomputing facilities in IMR, Tohoku Univ.

\newpage
\begin{figure}[h]
\begin{center}
\epsfig{file=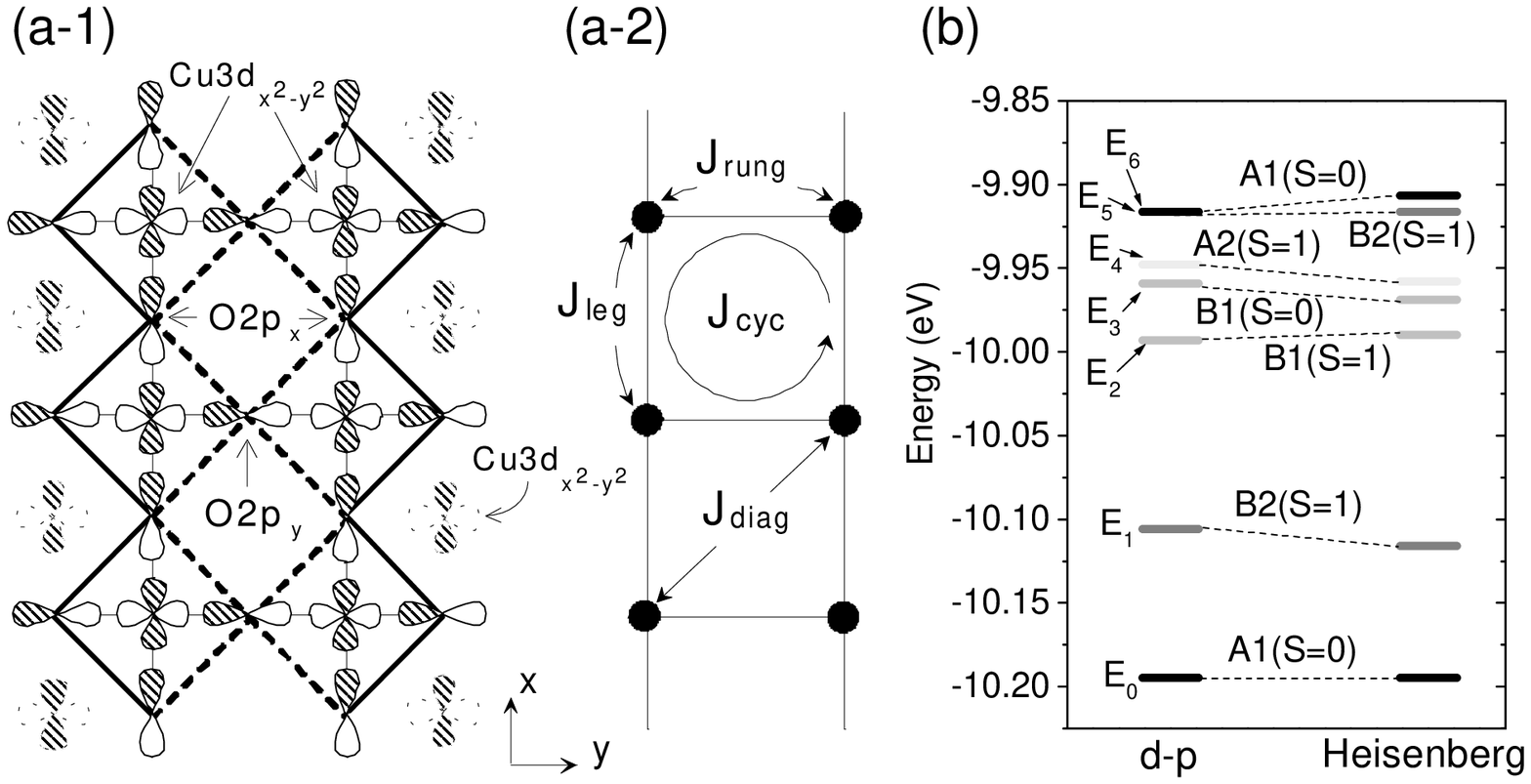,width=13cm,clip=}
\caption{(a-1) The $d$-$p$ model (Cu$_6$O$_{17}$ cluster) simulating a ladder cuprate. Unconnected orbitals denote Cu sites on the neighboring ladders. There are two inequivalent bonds for $T_{pp}$, which are distinguished by the solid and dotted lines. (a-2) The corresponding Heisenberg model. (b) The distribution of six lowest eigenstates for the $d$-$p$ model (a-1) and the Heisenberg model (a-2). Indices in figure denote the  irreducible representation and the total spin of the eigenstates.
}
\label{fig:1}
\end{center}
\end{figure}

\newpage
\begin{figure}[t]
\begin{center}
\epsfig{file=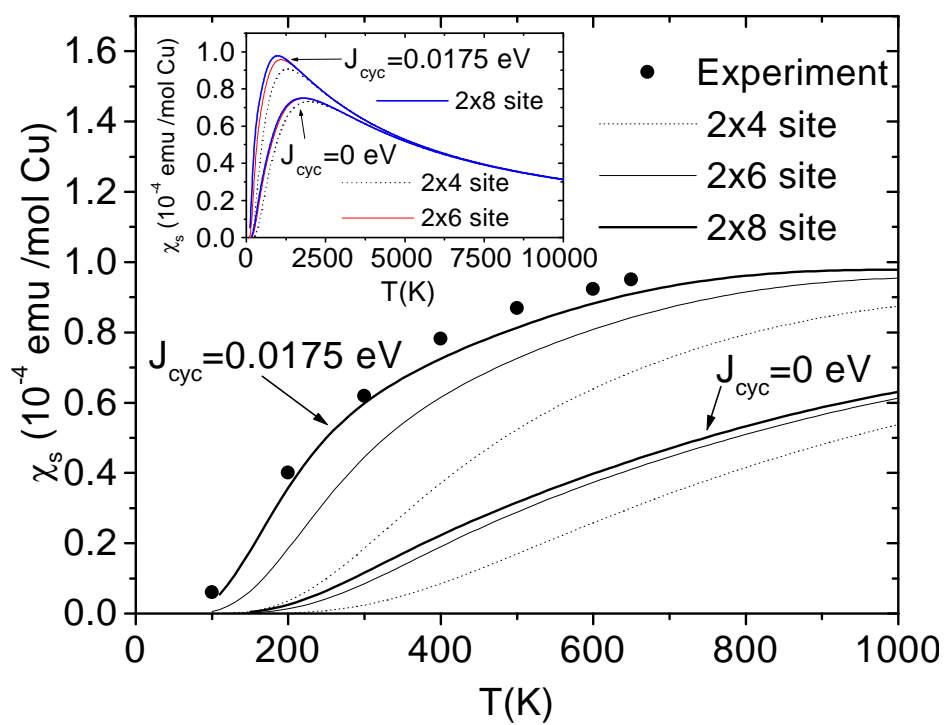,width=13cm,clip=}
\caption{The calculated $\chi(T)$'s for Heisenberg ladders. Experimental results are shown by filled circles.$^6$
%The results are shown for the parameters of $J_{\rm leg}$=0.19~eV, $J_{\rm rung}$=0.15~eV, $J_{\rm diag}$=0.003~eV and $J_{\rm cyc}$=0.0175~eV and of the same values but $J_{\rm cyc}$=0~eV.
}
\label{fig:3}
\end{center}
\end{figure}

\end{document}